# Optical and Acoustic Phonons in Turbostratic and Cubic Boron Nitride Thin Films on Diamond Substrates

Erick Guzman,[1] Fariborz Kargar,[1,*], Avani Patel,[2] Saurabh Vishwakarma,[3] Dylan Wright,[1,4], Richard B. Wilson,[5] David J. Smith,[3] Robert J. Nemanich,[2] and Alexander A. Balandin[1,4,*]

[1]Department of Electrical and Computer Engineering, University of California, Riverside, California 92521 USA

[2]Department of Physics, Arizona State University, Tempe, Arizona 85287 USA

[3]School of Engineering for Matter, Transport and Energy, Arizona State University, Tempe, Arizona 85287, USA

[4]Department of Materials Science and Engineering, University of California, Los Angeles California 90095 USA

[5]Department of Mechanical Engineering, University of California, Riverside, California 92521 USA

---

[*] Corresponding authors: fkargar@ece.ucr.edu (F.K.); balandin@seas.ucla.edu (A.A.B.)

# ABSTRACT

We report an investigation of the bulk optical, bulk acoustic, and surface acoustic phonons in thin films of turbostratic boron nitride (*t*-BN) and cubic boron nitride (*c*-BN) grown on B-doped polycrystalline and single-crystalline diamond (001) and (111) substrates. The characteristics of different types of phonons were determined using Raman and Brillouin-Mandelstam light scattering spectroscopies. The atomic structure of the films was determined using high-resolution transmission electron microscopy (HRTEM) and correlated with the Raman and Brillouin-Mandelstam spectroscopy data. The HRTEM analysis revealed that the cubic boron nitride thin films consisted of a mixture of *c*-BN and *t*-BN phases, with *c*-BN being the dominant phase. It was found that while visible Raman spectroscopy provided information for characterizing the *t*-BN phase, it faced challenges in differentiating the *c*-BN phase either due to the presence of high-density defects or the overlapping of the Raman features with those from the B-doped diamond substrates. In contrast, Brillouin-Mandelstam spectroscopy clearly distinguishes the bulk longitudinal and surface acoustic phonons of the *c*-BN thin films grown on diamond substrates. Additionally, the angle-dependent surface Brillouin-Mandelstam scattering data show the peaks associated with the Rayleigh surface acoustic waves, which have higher phase velocities in *c*-BN films on diamond (111) substrates. These findings provide valuable insights into the phonon characteristics of the *c*-BN and diamond interfaces and have important implications for the thermal management of electronic devices based on ultra-wide-band-gap materials.

**KEYWORDS:** ultrawide-bandgap materials; diamond; cubic boron nitride; Brillouin light scattering; Raman spectroscopy

1. **Introduction**

Ultra-wide bandgap (UWBG) semiconductors are attracting increasing attention owing to their potential for applications in high-power electronics and deep ultra-violet (UV) photonics [1–8]. Materials in the UWBG category have electronic bandgaps greater than that of GaN, typically > 4 eV [1,2]. Among these materials, *c*-BN has one of the highest breakdown fields projected to be more than 8 MV cm$^{-1}$ [9,10], a large electronic bandgap of ~6.4 eV [11], and high thermal conductivity of ~1600 Wm$^{-1}$K$^{-1}$, reaching 75% of that for diamond [12]. Only diamond outperforms *c*-BN in terms of electronic, thermal, and mechanical characteristics. Intrinsic UWBGs are electrical insulators and must be doped to be used in electronic devices. Diamond has limitations with regard to doping – it can be consistently *p*-doped, but achieving reliable *n*-type doping remains challenging [1]. In contrast, *c*-BN can be conveniently doped to become an *n*-type or *p*-type semiconductor [5,9,10]. The remaining critical issue with *c*-BN is the synthesis of single- or poly-crystalline thin films of phase-pure *c*-BN [13]. This task is difficult since the material favors the formation of the hexagonal sp$^2$-bonded phase, either in hexagonal, *h*-BN, or turbostratic, *t*-BN forms, depending on the growth process [14,15]. The in-plane crystal structure of the *t*-BN is similar to that of *h*-BN with the difference being loss of some or all order in the out-of-plane direction [16–18]. The *t*-BN planes are stacked and rotated in random directions along the *c*-axis and the interplanar distance becomes larger than that of *h*-BN. The planes may even have some tilt angle with respect to the plane normal [16–18]. The properties of the *t*-BN phase are less characterized compared to other BN phases and require in-depth investigations. There have been recent advances in the growth of *c*-BN thin films with high phase purity. Attempts so far have been directed towards the growth of *c*-BN on several substrates such as Si, Cu, Ni, and SiC/Si [15,19]. Growth of *c*-BN on different substrates typically requires a rather thick intermediate layer, such as *h*-BN or diamond, due to the lattice mismatch of the *c*-BN film with the underlying substrate [2]. An alternative approach is the synthesis of *c*-BN films directly on diamond substrates, considering the small lattice mismatch of only ~1.4% [20]. Recent studies have demonstrated that direct growth of *c*-BN on diamond substrates reduces the thickness of the *h*-BN intermediate phase, and in some cases, eliminates it entirely [20].

Another important advantage of the direct coupling of *c*-BN thin films with diamond substrates lies in their closely matched acoustic impedance, simple unit cell structures, vibrational similarity at the interface, and aligned phonon density of states [21,22]. These characteristics are crucial for efficient heat transport across adjoining interfaces which is defined by thermal boundary conductance [21]. Conventionally, the thermal boundary conductance at the interfaces is expressed using the acoustic mismatch model (AMM) or the diffuse mismatch model (DMM), respectively, in the low- and high-temperature ranges [23,24]. The acoustic impedance is defined as $\zeta = \rho v_s$ where $\rho$ and $v_s$ are the mass density and average phonon group velocity, respectively [24]. According to AMM, in heterostructures with ideal interfaces, the acoustic impedance mismatch of the constituent layers determines the thermal boundary conductance at the interface [24]. The smaller the acoustic mismatch of the constituents in layered structures, then the lower will be the temperature rise at the interface for a given heat flux. Based on AMM, since the acoustic mismatch of *c*-BN with diamond is negligibly small, heat transfers efficiently at the interface with little temperature rise from the *c*-BN layer to the underlying diamond substrate [22]. However, recent studies show that thermal boundary conductance is mostly affected by the bonding strength of constituent materials at the interface and the nature of the interfacial disorder. Owing to the small lattice mismatch of *c*-BN and diamond, one expects forming strong and highly structured interfaces that enhance heat transport across their interface. In fact, theoretical studies show that the interfacial thermal conductance at *c*-BN/diamond interface can reach ten times higher than that for the Si/diamond heterostructure [22]. In this sense, the premise of the direct coupling of diamond and *c*-BN layers, which are the two UWBG materials with the highest thermal conductivity, has important implications for the ongoing challenge of efficient thermal management of high-power UWBG-based electronic devices [25].

Acoustic phonons are the main heat carriers in both *c*-BN and diamond. The phonon properties of thin films can deviate significantly from theoretical predictions depending on the film phase purity, film/substrate interfacial structure, and atomic rearrangement within the film. Moreover, the phonon dispersion can change due to phonon confinement effects in thin film structures [26,27]. The latter results in the appearance of additional phonon modes other than the acoustic and optical phonon polarization branches in bulk materials. The contribution of these phonon modes to heat transport in thin film materials or at the interface of heterostructures is still a topic for debate [28,29]. From a theoretical perspective, the conventional AMM and DMM models typically consider only the average bulk phonon group velocity of each medium and do not account for possible effects of confined, guided, or interfacial phonon modes [30]. Recent studies have shown that interfacial phonon modes play an important role in heat conductance across the interfaces [29,31]. More importantly, the DMM fails to accurately predict the thermal conductance at the interface when two identical surfaces interface with each other [24,30]. In this case, one would expect the phonon transmission coefficient to approach one where the DMM predicts the phonon transmissivity to be 0.5 [30]. Given the negligible difference in the acoustic impedance and small lattice mismatch of *c*-BN and diamond, this pair of materials provides a convenient experimental system to investigate fundamental questions of heat transfer in thin films and across interfaces in low and high-temperature ranges. This task requires a thorough understanding of the bulk and surface phonon states of the *c*-BN films and their correlation with the film morphology, phase purity, and interlayer structure of the synthesized film on the diamond substrate.

Brillouin-Mandelstam light scattering spectroscopy (BMS) is a nondestructive optical technique that has been widely used for the detection of low-energy bulk and surface acoustic phonons in different material systems, including thin films [32–35]. This technique can determine the energy of bulk longitudinal acoustic (LA) and transverse acoustic (TA) phonons in the vicinity of the Brillouin-zone (BZ) center. The information obtained allows for the extraction of the phonon group velocity, *i.e.*, the sound velocity of LA and TA modes which can then be used to experimentally estimate the phonon transmissivity using the acoustic mismatch or diffuse mismatch models. Moreover, the method provides the additional benefit of obtaining the dispersion of surface phonon states close to the BZ center in the specific case of thin films or opaque materials [32–34]. Together with Raman spectroscopy, the BMS technique presents a comprehensive picture of the energy distribution of both optical and acoustic phonons close to the BZ center. To the best of our knowledge, there have been no previous studies of the bulk and surface phonon states of the *t*-BN and *c*-BN phases on diamond substrates.

In this work, we have used visible Raman and BMS spectroscopies to investigate the bulk optical and acoustic phonons, as well as surface acoustic phonons, in thin films of *t*-BN grown on B-doped polycrystalline diamond, and *c*-BN films grown on B-doped single-crystalline (001) and (111) diamond substrates. The thickness and the atomic structure of the samples were first characterized using high-resolution transmission electron microscopy (HRTEM). The TEM results were correlated with the data obtained from Raman and BMS experiments. The TEM analyses show that both boron nitride thin films grown on diamond substrates contain a mixture of *c*-BN and *t*-BN phases, with the *c*-BN being the dominant phase. Our results demonstrate that while visible Raman can be used to characterize the *t*-BN phase on diamond, it does not offer sufficient information for the *c*-BN phase either due to the high concentration of defects or the overlapping of its Raman features with that of the underlying B-doped diamond substrates. BMS, on the other hand, provides distinguishable peaks associated with the bulk longitudinal acoustic phonons of the *c*-BN phase on diamond substrates. The spectra of our angle-dependent surface Brillouin-Mandelstam scattering experiments on *c*-BN thin films exhibit peaks associated with the so-called Rayleigh surface acoustic waves. The phase velocity of Rayleigh modes is substantially higher for the *c*-BN samples on diamond (111) assessing its higher *c*-BN phase content, in agreement with the analyses acquired from HRTEM images. Our findings regarding the characteristics of surface and bulk phonons provide valuable insights into the thermal transport properties across *c*-BN and diamond interfaces, which have crucial implications for their applications in UWBG-based device applications.

2. Materials and Methods

The *t*-BN and *c*-BN films in this study were grown by the electron cyclotron resonance plasma-enhanced chemical vapor deposition (ECR PECVD) method [36] on various B-doped diamond substrates purchased from TISNCM and E6 vendors. The boron concentration in the substrates is estimated to be greater than $10^{20}$ cm$^{-3}$. Three heterostructure samples including a *t*-BN film on polycrystalline diamond, *c*-BN film on diamond (001), and *c*-BN film on diamond (111) were investigated in this study. For simplicity, these samples are referred to as *t*-BN, *c*-BN-1, and *c*-BN-2, respectively. The *t*-BN film on the B-doped polycrystalline diamond substrate was grown with a gas flow ratio of H$_2$:BF$_3$ of 4:4 (gas flow ratio in standard cubic centimeters per minute or sccm) at a temperature of 735 °C for 2 hours and 30 min. The *c*-BN-1 film was grown with a ratio of $H_2:BF_3$ of 2:2 sccm at a temperature of 675 °C for 2 hours and 30 min. The *c*-BN-2 film was prepared with an $H_2:BF_3$ ratio of 1.5:2 for 44 min and then with an $H_2:BF_3$ ratio of 2:2 sccm for 2 hours and 16 min at a temperature of 735 °C for a total of 3 hours. The details of the growth process have been reported elsewhere [36].

Samples suitable for TEM observation were prepared using focused-ion-beam milling with a Thermo Fisher Helios G5-UX dual-beam system. Cross-sectional TEM images were taken using a Phillips-FEI CM-200F operated at 200 kV and an FEI Titan 80-300 operated at 300 kV.

Raman spectroscopy was conducted in the conventional backscattering configuration using a blue laser with an excitation wavelength of 488 nm. The power on the samples was 5 mW during all measurements. The laser spot diameter on the sample was ~1 μm. BMS experiments were performed in the backscattering geometry using a continuous-wave solid-state diode-pumped laser operating at the excitation wavelength of $\lambda = 532$ nm. The incident light was *p*-polarized whereas the polarization of the scattered light was not analyzed. The laser light was focused using a lens with a numerical aperture of NA=0.24. The scattered light was collected using the same lens and directed to the high-resolution high-contrast 3+3 tandem Fabry-Perot interferometer (The Table Stable, Switzerland), detector, and spectrum analyzer. To probe surface acoustic waves, samples were rotated so the angle of the incident light can be varied between 30° to 70° with respect to the sample's surface normal.

To interpret the BMS data, one needs to determine the index of refraction, $n$, of the thin films at BMS laser excitation wavelength. Measurements of the refractive index were made using a Filmetrics F40-UV microscope-based spectrometer. Light with wavelengths in the range of 190 – 1100 nm was used to measure the sample reflectance. The average spot size of the light on the sample was ~35 μm. The thickness and optical constants of the thin films were extracted by fitting the reflectance data using the governing equations.

## 3. Results and Discussion

3.1. TEM analysis

Figures 1 (a,b) show TEM images of the *t*-BN thin film on the B-doped polycrystalline diamond at different magnifications. Figure 1 (b) exhibits the atomic configuration of the *t*-BN film in more detail. As visible in most regions, the sp$^2$-bonded *h*-BN sheets of *t*-BN grew close to the vertical direction with their *c*-axis roughly parallel to the sample surface (see the yellow traces in Figure 1 (b)). However, in a few regions, the *t*-BN planes are also tilted (see the red traces in Figure 1 (b)). The total thickness of the *t*-BN film was estimated to be in the range of $240 \pm 10$ nm.

Figures 1 (c,d) present TEM images of the *c*-BN-1 sample. The dark line at the center of Figure 1 (c) is the interface between the film and the diamond substrate which contains significant hydrogen defects due to the substrate cleaning procedure using $H_2$-plasma. It is estimated that the hydrogen is penetrated as much as ~10 nm into the diamond substrate thus reducing the interface quality. As visible in Figure 1 (d), the boron nitride film in this sample is a mixture of *t*-BN and *c*-BN phases. The *c*-BN regions, depicted by yellow rectangles, are the lighter clustered regions, while the *t*-BN spots, shown by red rectangles, look like clear parallel lines with various tilt angles. The *t*-BN phase is predominant at the substrate/interface and in the intermediate regions. The top surface of the film is terminated with mostly the *c*-BN phase, although the *t*-BN phase was also detected in some regions. The *c*-BN regions are oriented randomly with grain sizes of ~5-10 nm. The total thickness of the film was estimated to be $168 \pm 18$ nm.

The structure and morphology of the *c*-BN-2 film are shown in Figures 1 (e,f). This sample has the higher phase purity of the *c*-BN regions with no signs of the *t*-BN phase at the top surface. The *c*-BN and *t*-BN regions are depicted by yellow and red rectangles in Figure 1 (f). A careful analysis of the film interface with the underlying diamond (111) substrate confirms the existence of a mixture of *h*-BN, *w*-BN, and *t*-BN phases with varying thicknesses. The film is predominantly constituted of *c*-BN grains, randomly oriented in different crystallographic planes. The total thickness of the film is $157 \pm 6$ nm. This film is essentially epitaxially oriented, phase pure c-BN at the diamond interface. Towards the surface, small t-BN domains are evident, and the c-BN regions show increased orientational randomness.

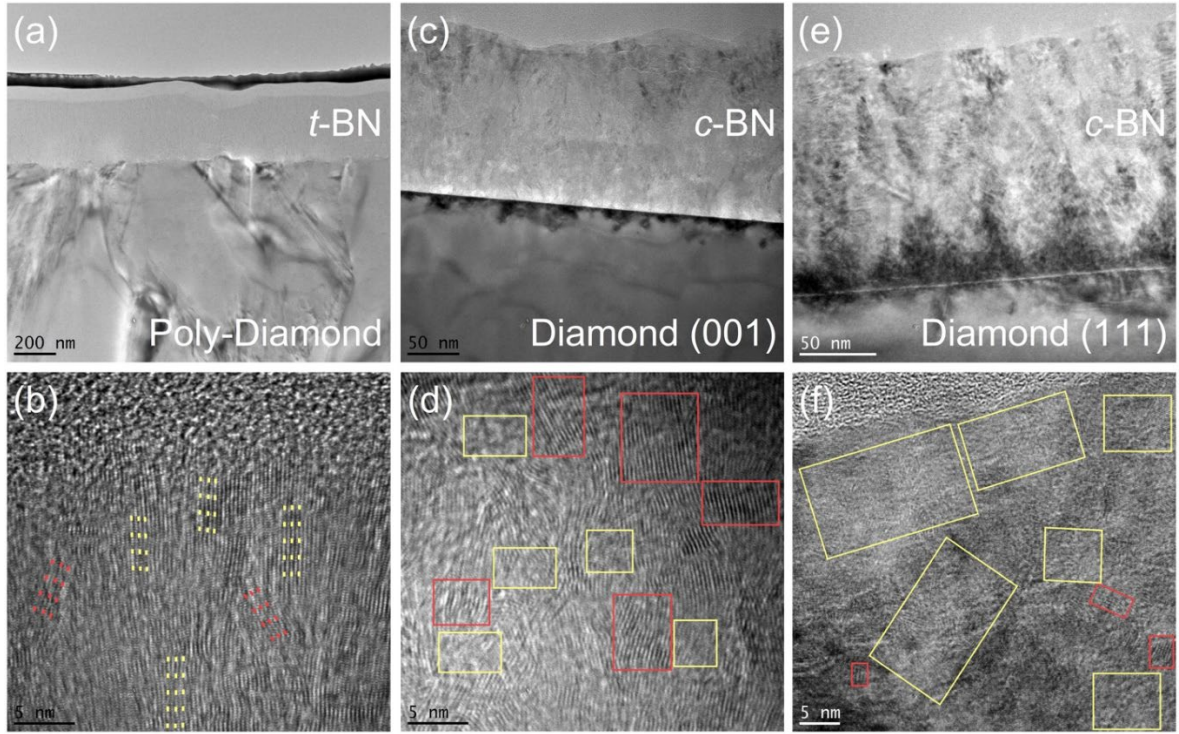

Figure 1: TEM images of boron nitride thin films grown on diamond substrates. (a) The *t*-BN film on the B-doped polycrystalline diamond. (b) Higher magnification image of the *t*-BN film showing the arrangement of the BN sheets. The yellow traces exhibit the vertically aligned sp$^2$-bonded BN sheets whereas the red traces show some regions where the BN planes are tilted slightly. (c)The *c*-BN-1 film, interface, and the underlying (001) single-crystal diamond substrate. The dark line shows the interface of the film and the substrate, damaged by hydrogen. (d) Enlarged image of the BN film structure confirming the presence of a mixture of *t*-BN and *c*-BN phases, surrounded by red and yellow rectangles, respectively. (e) The *c*-BN-2 film grown on a single crystal (111) diamond. (f) High-resolution image of the *c*-BN-2 sample confirming the dominant *c*-BN phase (yellow zones) compared to the *t*-BN phase (red zones). The *c*-BN regions near the surface show increased orientation randomness.

3.1 Raman analysis

A series of Raman experiments were conducted on the BN thin films and their respective underlying substrates. To cross-check the results obtained, the same experiments were carried out on bulk single-crystal *c*-BN as the reference sample. Figure 2 (a) presents the Raman spectra of the *t*-BN film (red line) and its underlying B-doped polycrystalline substrate (black line). The sharp peak observed at ~1330 cm$^{-1}$ in both spectra is associated with the zone center Raman peak (ZCP) of the diamond substrate. The ZCP of the undoped diamond is located at 1332 cm$^{-1}$ which in our samples downshifted by 2 cm$^{-1}$ which is attributed to to the phonon softening as a result of the heavy boron doping and clustering [37–39]. The presence of this peak in both spectra confirms that a significant fraction of the incident light interacts with the diamond substrate even though the light is focused on the top surface of the *t*-BN layer. Below the ZCP, the pristine B-doped polycrystalline diamond exhibit broad features at ~607 cm$^{-1}$, 897 cm$^{-1}$, and 1042 cm$^{-1}$. These three peaks are the Raman-forbidden bands of diamond which are activated due to the presence of boron dopants [40–42]. The spectrum of the *t*-BN film on the B-doped polycrystalline diamond exhibits quite different features (see the red curve in Figure 2 (a)). Except for the strong ZCP, other Raman features, present in the substrate, are not visible in the *t*-BN spectrum, but new Raman peaks located at 504 cm$^{-1}$, 1220 cm$^{-1}$, 1370 cm$^{-1}$, 1538 cm$^{-1}$, and 2803 cm$^{-1}$ are detected. The weak feature at 1370 cm$^{-1}$ is attributed to the *t*-BN phase [17,18] which is almost hidden due to the sharp ZCP feature and the high background signal from the substrate. The nature of this peak can be thought of as an upshifted single-crystal *h*-BN peak which is expected to be observed at ~1364 cm$^{-1}$ [43–45]. The peaks at 1220 cm$^{-1}$ and 1538 cm$^{-1}$ most likely originate from the maximum phonon density of states of *t*-BN, assuming its crystal similarity with *h*-BN [45]. Other peaks are all associated with the B-doped diamond substrate. The peak at 504 cm$^{-1}$ is likely the same as the 607 cm$^{-1}$ peak observed in the substrate. It is redshifted to lower wavenumbers presumably due boron clustering in the heavily B-doped substrate. This peak is associated with the maximum phonon density of states of diamond and is reported to downshift as the boron concentration increases [40]. The broad peak found at ~2803 cm$^{-1}$ has been observed previously in Raman of *h*-BN [45], but no obvious assignment of this peak has been reported or could be made here.

Now, we investigate the Raman characteristics of bulk single-crystal *c*-BN and *c*-BN thin films on diamond substrates. The bulk single-crystal c-BN samples were outsourced from Hyperion Materials and Technology, the USA. Samples were in powder form with average size of ~300 μm to ~600 μm. The Raman spectrum of the polycrystalline *c*-BN powder is presented in Figure 2 (b). The spectrum shows two peaks attributed to the TO and LO modes, respectively, at 1054 cm$^{-1}$ and 1304 cm$^{-1}$. The spectral position of these peaks is in excellent agreement with previously reported values [45–47]. Figure 2 (c) shows the results of Raman measurements on the *c*-BN-1 thin film (red curve), and the B-doped single-crystal diamond (001) substrate without *c*-BN film (black curve). Interestingly, no signatures of the TO and LO peaks of *c*-BN were detected in this sample. One may assume that these peaks may have been buried under the intense Raman features coming from the substrate. Given that the TO Raman peak of *c*-BN at 1054 cm$^{-1}$ should be intense (see Figure 2(b)) and there are no significant diamond features at that wavenumber, this seems less likely. An alternative explanation is that the TO and LO Raman bands in *c*-BN are not detectable due to the presence of high-density defects and the consequent relaxation of Raman selection rules. The latter causes weak scattering from optical phonon modes with different wavevectors. This explanation seems more in line with the characteristics of our samples. The only difference in the two spectra obtained from the film and substrate is the presence of a weak Raman feature at 1135 cm$^{-1}$ that could be related to the second-order Raman scattering from the zone-edge LO(L) or LO(X) modes [45]. Figure 2 (d) presents the Raman spectra of the *c*-BN-2 thin film grown on the single-crystal diamond (111) substrate. The two spectra are almost identical, and no peaks associated with the *c*-BN layer could be detected.

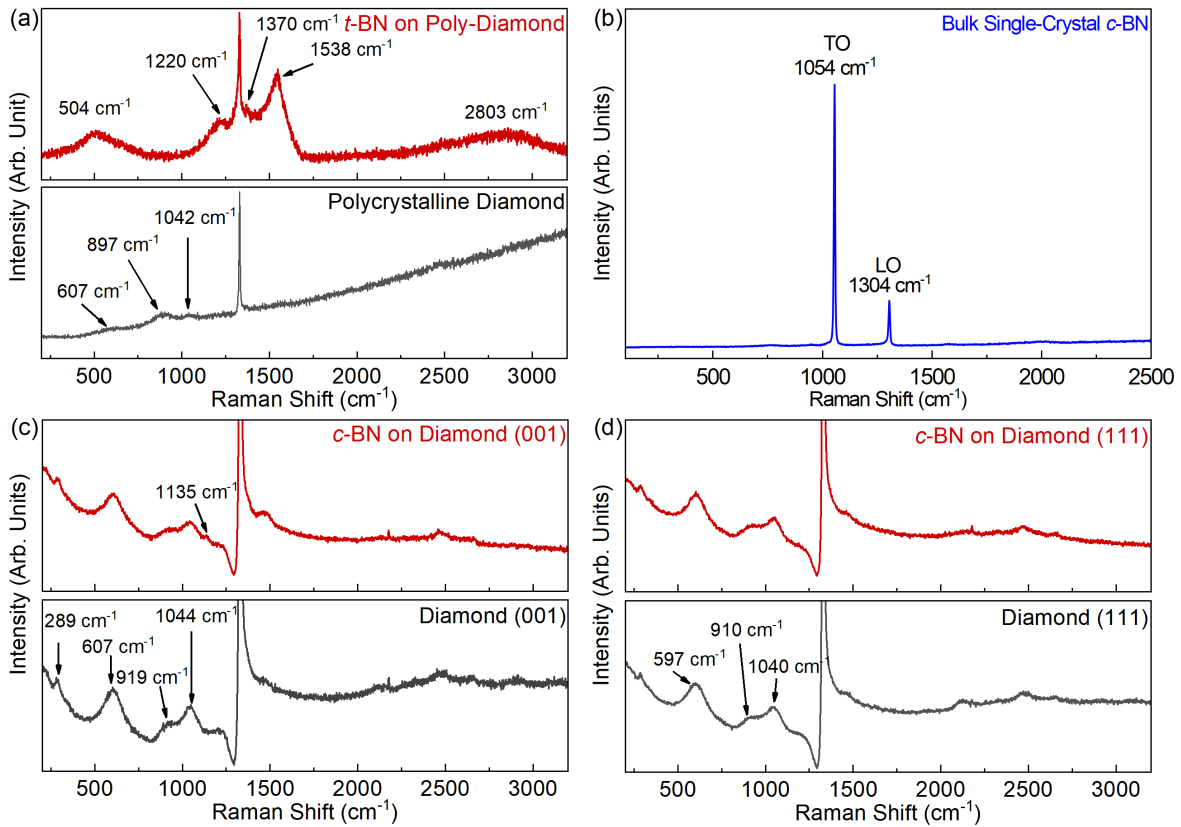

Figure 2: Raman spectra of the BN samples and their respective B-doped diamond substrates under a 488 nm excitation laser. (a) *t*-BN film on B-doped polycrystalline diamond, (b) bulk single-crystal *c*-BN, (c) *c*-BN film on B-doped single-crystal diamond (001), and (d) *c*-BN film on B-doped single-crystal diamond (111).

3.2 BMS results

We now turn to the main element of this work – the observation of acoustic phonons in boron nitride films using the BMS technique. Figure 3 shows the results of the BMS measurements of bulk phonons in the single crystal *c*-BN and the three BN thin films in the frequency range of 25 GHz to 175 GHz. Symmetric peaks on each side of the spectra are associated with the Stokes and Anti-Stokes scattering processes. The spectral positions of the observed peaks, $f$, were determined accurately by fitting the experimental data with individual Lorentzian functions. In BMS, the phonon wave vector of bulk phonon modes that contribute to light scattering is $q_B = 4\pi n/\lambda$, where $\lambda = 532$ nm is the laser excitation wavelength and $n$ is the refractive index of the medium at $\lambda$ [32]. The refractive indices of *t*-BN on polycrystalline diamond and *c*-BN on diamond (001) samples could not be measured due to their rough surfaces. To calculate the BMS phonon wavevector, we used the previously reported values of $n$ for the *t*-BN and *c*-BN-1 as 1.870 and 2.117, respectively [48]. Note that one should use these values with caution to calculate the BMS phonon wave vector. *t*-BN is a disordered form of *h*-BN along the *c*-axis stacking direction. *h*-BN sheets are optically anisotropic and given that in our *t*-BN film, its constituent *h*-BN sheets are lined up in different orientations close to vertical, taking the reported value for the index of refraction could impose significant errors. The *c*-BN film on diamond (001) also contains mixed phases of *c*-BN and *t*-BN. The phase proportions in these samples could not be identified from TEM images, and the correct value of $n$ thus remains unclear. The refractive index of the *c*-BN on diamond (111) substrate was measured as $n = 2.12 \pm 0.21$ which agrees well with the available data [48,49].

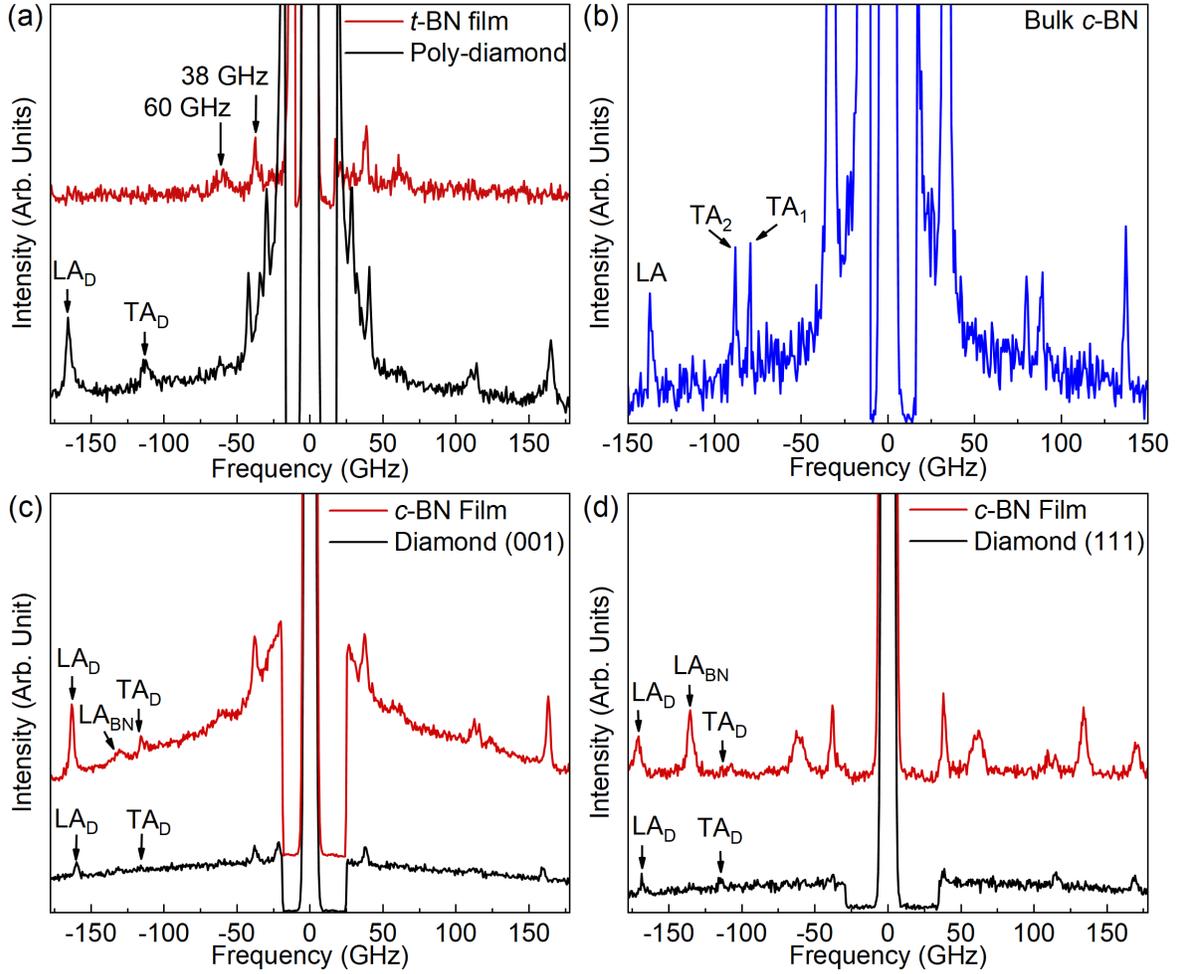

Figure 3: Brillouin light scattering spectra of bulk phonon modes in (a) *t*-BN thin film on B-doped polycrystalline diamond, (b) bulk single-crystal *c*-BN, (c) *c*-BN thin film on diamond (001), and (d) *c*-BN thin film on diamond (111). The peaks labeled as "LA" and "TA" correspond to the longitudinal and transverse acoustic bulk phonons, respectively. The subscripts "D" and "BN" represent "diamond" and "boron nitride" in all panels.

Figure 3 (a) exhibits the BMS spectra of the polycrystalline diamond substrate (black line) and the *t*-BN thin film (red line). Two peaks labeled as "$LA_D$" and "$TA_D$" at $f \sim 165$ GHz and $f \sim 112$ GHz, correspond to the longitudinal and transverse acoustic phonons, respectively, of the polycrystalline diamond. The spectral positions of the LA and TA peaks agree well with the previously reported data [50]. The *t*-BN thin film exhibits two distinct broad peaks at 38 GHz and 60 GHz. The peak at 60 GHz is detectable only in the *p-p* or *s-s* incident-scattered light polarization configuration which is a characteristic of the bulk LA phonons in BMS experiments. The peak at 38 GHz is visible in the cross-polarization configuration (*p-s*), a property of TA phonons (see Supplementary

Information). Thereby, we attribute these peaks to *t*-BN's transverse acoustic and longitudinal acoustic phonon branches, respectively. Assuming the refractive index of $n = 1.870$, one can extract the phase velocity of these phonon branches according to $v = 2\pi f/q_B = \lambda f/2n$. It should be noted that in the region near the center of the Brillouin Zone (BZ), the dispersion of acoustic phonons follows a linear relationship. Thus, the phase velocity of these phonon branches is equal to their group velocity or "sound velocity". The calculated phonon group velocities for the TA and LA phonons in the *t*-BN film are $v_{TA} = 5,405$ ms$^{-1}$ and $v_{LA} = 8,535$ ms$^{-1}$, respectively. It is worthwhile comparing the spectral positions of these peaks with those reported for *h*-BN. The significance of comparing *t*-BN results with those for *h*-BN lies in the fact that as stated previously, the *t*-BN phase constitutes from tilted disorder *h*-BN sheets with slightly increased interlayer distance [16–18]. The previous reports on the Brillouin light spectroscopy measurements of bulk *h*-BN show LA phonon peaks at ~28 GHz and ~112 GHz with phonon wave vector along and perpendicular to the *c*-axis of *h*-BN, respectively [51]. This is expected due to the large anisotropy in elastic properties of *h*-BN, with a slow LA phonon velocity of ~3,279 ms$^{-1}$ along the *c*-axis compared to a fast LA phonon velocity of 18,595 ms$^{-1}$ perpendicular to the *c*-axis. In our experiments, the examined phonon wave vector was close to the surface normal of the film. Given most of its *h*-BN sheets in the *t*-BN film are aligned vertically (see Figure 1 (a)), it is reasonable to observe LA phonon velocity at some weighted average frequency of LA phonons propagating along and perpendicular to *h*-BN *c*-axis.

To characterize the acoustic phonon properties of *c*-BN films, BMS experiments were first conducted on bulk single-crystal *c*-BN as the reference sample. Figure 3 (b) shows the accumulated BMS spectra. Three sharp peaks at 79.6 GHz, 88.1 GHz, and 137.1 GHz are attributed to two TA and one LA phonon modes. The Brillouin phonon wave vector in these experiments is $q_B = 50.01$ μm$^{-1}$, assuming $n = 2.117$ at $\lambda = 532$ nm for *c*-BN [48,49]. The corresponding phonon group velocities for LA and TA phonons in single-crystal *c*-BN are calculated as 17,225 ms$^{-1}$, 11,069 ms$^{-1}$, and 10,000 ms$^{-1}$, respectively. The obtained values for the bulk *c*-BN were used as a reference for the characteristics of *c*-BN thin films on diamond substrates.

Figure 3 (c) shows the spectra for the c-BN thin film on diamond (001). The black and red curves are the spectra for the pristine diamond and the thin film, respectively. As seen, in the diamond substrate, two peaks were detected at 116.6 GHz and 159.4 GHz, corresponding to the TA and LA phonon branches along the diamond [100] crystallographic direction. The BMS spectrum for the c-BN shows an additional broad and weak peak located at 129.2 GHz. This peak is attributed to the LA$_{BN}$ phonon of c-BN. The spectral position of this peak is ~8 GHz lower than that for the LA peak of the bulk single-crystal c-BN, presumably due to the existence of some t-BN regions mixed with the dominant c-BN phase. Assuming $n = 2.117$, one would extract a phonon group velocity of $v_{LA} = 16,234$ ms$^{-1}$, almost 6% lower than that for the single-crystal c-BN sample. The presence of a small fraction of t-BN mixed phase manifests itself in a slower phonon group velocity compared to that of the pure phase c-BN. Figure 3 (d) presents the BMS spectra for the c-BN film on a diamond (111) substrate. The TA$_D$ and LA$_D$ peaks for the diamond with phonon wave vector along [111] direction appear at 114.4 GHz and 168.5 GHz, respectively (black curve). The spectrum for the c-BN film exhibits a sharp and well-defined Lorentzian peak at 135.4 GHz which is attributed to the LA$_{BN}$ phonon branch. As seen, the spectral position of this peak is only ~1.7 GHz lower than the single-crystal bulk c-BN sample. Using the measured refractive index of $n = 2.12$, the phonon group velocity is calculated to be $v_{LA} = 16,989$ ms$^{-1}$ which is only 1.4% lower than that for the single-crystal bulk c-BN. The latter confirms that the c-BN on diamond (111) sample has a higher order of c-BN phase purity. These findings align with the TEM results indicating higher content of c-BN phase in thin films grown on diamond (111) substrate compared to that of the diamond (001). Other peaks that appeared at much lower frequencies are associated with the scattering of light by surface phonons, or surface acoustic waves, which are discussed in the following.

In our experiments, the BMS spectra revealed several peaks in the low-frequency range. These features are related to the scattering of light by surface ripples created by surface acoustic waves propagating along the film layer. The phonon wave vector associated with these surface phonons in the backscattering geometry is $q_s = 4\pi \sin(\theta)/\lambda$ in which $\theta$ is the incident angle and $\lambda$ is the laser excitation wavelength [32–34]. The direction of the surface phonon wave vector lies in the scattering plane parallel to the film surface plane. The magnitude of $q_s$ only depends on the incident angle, $\theta$, and is no longer a function of the refractive index, $n$. Therefore, one can obtain the dispersion of surface phonons, *i.e.*, $f$ as a function of $q_s$, by changing the incident angle. Note that the types of surface acoustic phonons which can be detected via BMS in the supported thin films strongly depend on the transparency, thickness, crystal structure, relative acoustic velocity of the film with respect to its underlying substrate, and its displacement vibrational profile [52]. Cubic BN is among the materials with extraordinarily high acoustic phonon velocities. However, compared to diamond, acoustic phonons travel at slower velocities. In the case of slow films on fast substrates, discrete phonon modes associated with Rayleigh and Sezawa waves may appear in the BMS spectra, depending on the thickness of the films [52,53].

The surface Brillouin light scattering experiments were restricted to the *c*-BN thin films on diamond substrates. The surface of the *t*-BN sample was rather rough making it impossible to accumulate data within a reasonable accumulation time. The incident angle, $\theta$, was varied to select the $q_s$, and the spectral position of the BMS peaks were plotted as a function of $q_s$ to obtain the phonon dispersion. The phase velocity of the peaks was extracted according to $v_s = \omega/q_s = 2\pi f/q_s$. Note that surface phonons in layered structures are dispersive, *i.e.*, the phase velocity of phonons is no longer equal to their group velocity. Figures 4 (a,b) present the results of the surface Brillouin scattering measurements performed on the *c*-BN thin film grown on diamond (001) and diamond (111), respectively. The raw data is presented as scattered dots and the peaks were fitted using individual Lorentzian functions (solid lines). The *c*-BN film on diamond (001) exhibits two peaks, marked as R$_1$ and R$_2$, which are attributed to the first and higher-order Rayleigh waves, respectively. The frequency of these peaks increases with increasing $\theta$ and thus, $q_s$. In the case of *c*-BN on diamond (111) presented in Figure 4 (b), the data shows two peaks as well. The spectral position of the peak at the lower frequency range changes with varying $q_s$. This peak is denoted as $R_1$ and is associated with the Rayleigh acoustic wave. However, the frequency of the peak labeled with an '*' and located at 38 GHz remains constant. The origin of this 'spurious' peak is unclear. The peak at 38 GHz was occasionally observed in the data presented for the diamond substrate as well, and it is possible that this peak originated from the substrate as *c*-BN is transparent in the visible light region. Since the frequency of the Rayleigh peaks in thin film structures depend on the film thickness, it is conventional to present the spectral position of the surface acoustic Rayleigh waves as a function of the nondimensional parameter $q_s h$ (Figure 4 (c)).

The average thickness of the samples, $h$, for the $c$-BN-1 and $c$-BN-2 samples are $168 \pm 18$ nm and $157 \pm 6$ nm. These values are based on measurements of the TEM images shown in Figure 1.

Figure 4 (d) shows the calculated phase velocity as a function of $q_s h$. As expected, the phase velocity of the Rayleigh waves decreases with increasing $q_s h$ confirming its dispersive nature in thin film structures. Notably, the phase velocity of these modes is more than two times higher in the $c$-BN thin film on diamond (111), compared to that of the film on the diamond (001) substrate. The dispersion and velocity of Rayleigh waves in thin film structures depend on the thickness, the elastic coefficients of the substrate and the film, and the quality of bonding at the interface [54,55]. Due to the dissimilar crystallographic directions of the substrates used and the different thicknesses of the films, a reasonable comparison of the Rayleigh wave characteristics in these samples is not feasible. However, we did not expect to observe such a significant disparity in phase velocities for first-order Rayleigh waves (R1, in Figure 4 (d)) in two samples. The latter is most likely attributed to the better phase purity and interface quality of the films grown on diamond (111) sample, which is in agreement with the TEM measurements.

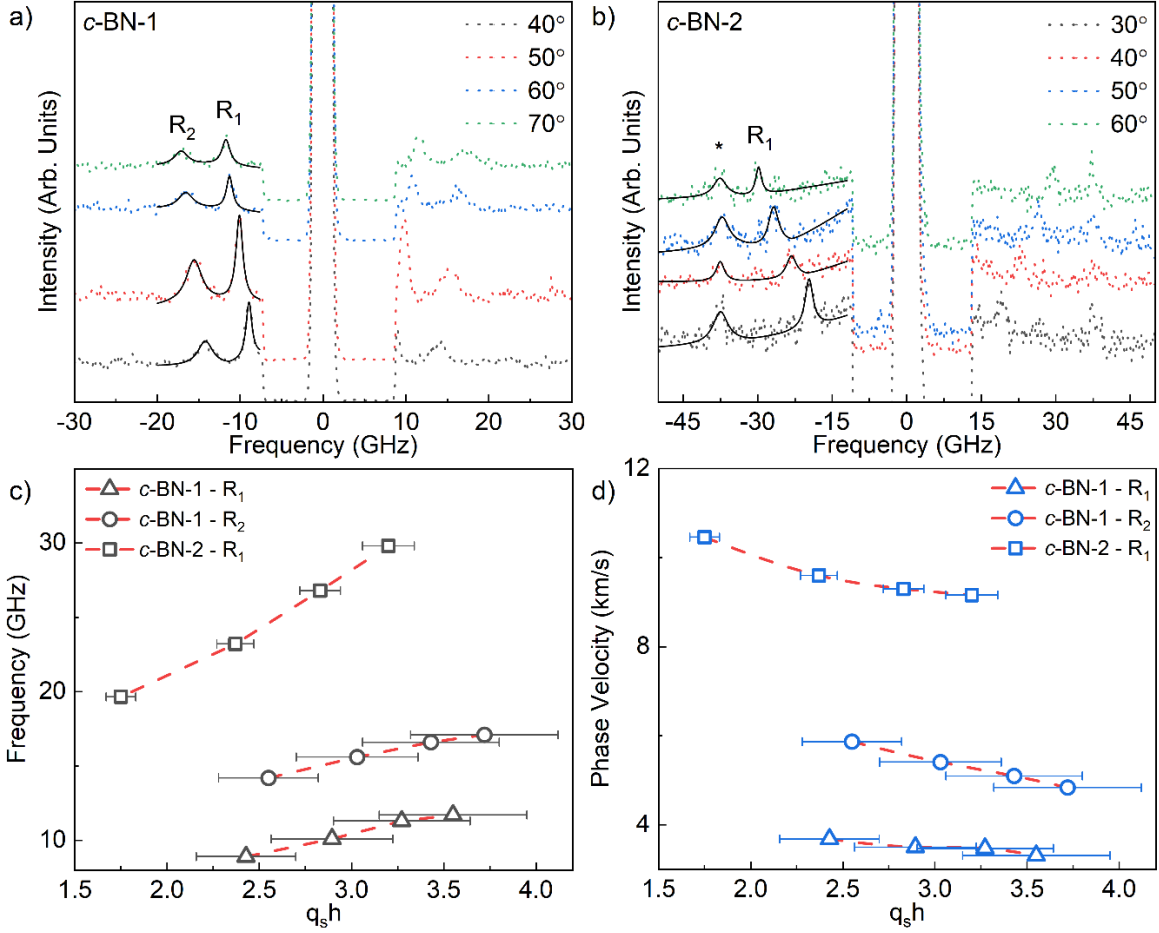

Figure 4: Brillouin spectra of the *c*-BN thin films: (a) diamond (001); and (b) diamond (111) substrates. Peaks denoted by "R" represent Rayleigh acoustic waves. The nature of the peak labeled with an "*" is unclear. Spectral position and the calculated phase velocity of the surface phonons of *c*-BN films: (c) diamond (001); and (d) diamond (111) substrates. Raw data in (a) and (b) is presented as dotted lines and the fitted data is shown as solid black lines.

To estimate the thermal interface conductance of *c*-BN on diamond samples, it is worth calculating the phonon transmissivity across the interfaces based on the BMS results using the acoustic mismatch model. The TA peaks could not be observed in the BMS measurements of *c*-BN thin films and thus, the analysis is limited to the calculation transmissivity coefficient of LA phonon branches. According to AMM, at normal incidence, the transmission coefficient from the film to the substrate is described as, $\tau_{f \to s} = 1 - R_{f \to s}$, in which the reflectivity, $R_{f \to s}$ is a function of acoustic impedances of the two media, $R_{f \to s} = |(\zeta_f - \zeta_s)/(\zeta_f + \zeta_s)|^2$. Table 1 summarizes the experimental bulk phonon velocities of the films and their respective substrates obtained from BMS experiments and the phonon transmission coefficient for the LA branch at the interface. As seen, the transmissivity coefficient for the *c*-BN on diamond samples reaches almost unity demonstrating that the thermal boundary resistance in these heterostructures could be among the lowest in UWBG heterostructures. The latter has important implications in the efficient thermal management of high-power UWBG-based heterostructure devices in which effective heat dissipation remains a formidable challenge.

**Table 1: Phonon group velocity and transmission coefficient at interfaces**

| Sample | $n$ | $\rho$ [kgm$^{-3}$] | $v_{TA1}$ [ms$^{-1}$] | $v_{TA2}$ [ms$^{-1}$] | $v_{LA}$ [ms$^{-1}$] | $\zeta$ [MPa.s.m$^{-3}$] | $\tau_{f \to s}$ (%) |
|---|---|---|---|---|---|---|---|
| Bulk single-crystal *c*-BN | 2.117 | 3450 | 10,002 | 11,070 | 17,227 | 59.43 | NA |
| *t*-BN | 1.87 | 2100 | 5,405 | | 8,535 | 17.92 | 68.5 |
| Polycrystalline diamond | 2.43 | 3530 | 12,260 | | 18,062 | 63.76 | |
| *c*-BN-1 | 2.117 | 3450 | | | 16,234 | 56.00 | 99.7 |
| Diamond (001) | 2.43 | 3530 | 12,764 | | 17,449 | 61.59 | |
| *c*-BN-2 | 2.117 | 3450 | | | 16,989 | 58.61 | 99.7 |
| Diamond (111) | 2.43 | 3530 | 12,523 | | 18,445 | 65.11 | |

## 4. Conclusions

The bulk and surface phonons in *t*-BN and *c*-BN thin films grown on diamond substrates have been investigated using visible Raman and Brillouin-Mandelstam light scattering spectroscopy. The results were correlated with the phase purity and film-substrate interface quality analyses of samples obtained from the corresponding high-resolution transmission electron microscopy images of the same films. Raman characterization did not provide sufficient information about the *c*-BN thin films; either due to the presence of defects or because of strong Raman features of the underlying B-doped diamond substrates. In contrast, the bulk and surface BMS measurement revealed well-distinguished peaks associated with the BN thin films. It was found that bulk acoustic phonons appeared at higher energies in BMS spectra for the *c*-BN thin film grown on diamond (111) compared to that of the film grown on diamond (001). The spectral position of the longitudinal acoustic phonons in the *c*-BN film on diamond (111) was close to that of the bulk single-crystal *c*-BN confirming its higher *c*-BN phase purity. Additionally, the results of the surface Brillouin scattering experiments revealed more than two times faster phase velocity for Rayleigh acoustic waves in the *c*-BN film on diamond (111) sample, which can be attributed to its higher phase purity and better film-interface quality. These findings agree well with the results of our TEM analyses. Our findings attest to the BMS as a powerful tool in the characterization of phonon properties in ultra-wide bandgap heterostructures. The information about the acoustic bulk and surface phonon properties in *c*-BN thin films has important implications in evaluating the thermal interfacial conductance in layered structures.

**ACKNOWLEDGEMENTS**

This work was supported by Ultra Materials for a Resilient, Smart Electricity Grid (ULTRA), an Energy Frontier Research Center (EFRC) funded by the U.S. Department of Energy, Office of Science, Basic Energy Sciences under Award # DE-SC0021230. A.A.B. and F.K. acknowledge the support of the National Science Foundation (NSF) via a Major Research Instrumentation (MRI) project DMR 2019056 entitled "Development of a Cryogenic Integrated Micro-Raman-Brillouin-Mandelstam Spectrometer." The authors also acknowledge the use of facilities within the John M. Cowley Center for High Resolution Electron Microscopy at Arizona State University, supported in part by the NSF Program No. NNCI-ECCS-1542160. The authors thank Dr. Menno Bouman and Dr. Jake Polster for the refractive index measurements at the KLA Instruments group.

**CONTRIBUTIONS**

A.A.B. and F.K. conceived the idea, coordinated the project, contributed to experimental data analysis, and led the manuscript preparation; E.G. conducted Raman and BMS experiments and contributed to the data analysis; D.W. contributed to the refractive index measurements; F.K. supervised the BMS and Raman experiments; A.P. prepared the $t$-BN and $c$-BN thin films; S.V. conducted TEM measurements; D.J.S. supervised the TEM measurements and analyses; R.W. assisted with the data analyses; R.J.N. provided the samples and supervised the BN growth; All authors contributed to the manuscript preparation.

**REFERENCES**

[1] J.Y. Tsao, S. Chowdhury, M.A. Hollis, D. Jena, N.M. Johnson, K.A. Jones, R.J. Kaplar, S. Rajan, C.G. Van de Walle, E. Bellotti, C.L. Chua, R. Collazo, M.E. Coltrin, J.A. Cooper, K.R. Evans, S. Graham, T.A. Grotjohn, E.R. Heller, M. Higashiwaki, M.S. Islam, P.W. Juodawlkis, M.A. Khan, A.D. Koehler, J.H. Leach, U.K. Mishra, R.J. Nemanich, R.C.N. Pilawa-Podgurski, J.B. Shealy, Z. Sitar, M.J. Tadjer, A.F. Witulski, M. Wraback, J.A. Simmons, Ultrawide-Bandgap Semiconductors: Research Opportunities and Challenges, Adv Electron Mater. 4 (2018) 1600501.

[2] M.H. Wong, O. Bierwagen, R.J. Kaplar, H. Umezawa, Ultrawide-bandgap semiconductors: An overview, J Mater Res. 36 (2021) 4601–4615.

[3] A. Soltani, H.A. Barkad, M. Mattalah, B. Benbakhti, J.C. De Jaeger, Y.M. Chong, Y.S. Zou, W.J. Zhang, S.T. Lee, A. BenMoussa, B. Giordanengo, J.F. Hochedez, 193 nm deep-ultraviolet solar-blind cubic boron nitride based photodetectors, Appl Phys Lett. 92 (2008) 53501.

[4] Y. Zou, Y. Zhang, Y. Hu, H. Gu, Ultraviolet detectors based on wide bandgap semiconductor nanowire: A review, Sensors. 18 (2018) 2072.

[5] S. Noor Mohammad, Electrical characteristics of thin film cubic boron nitride, Solid State Electron. 46 (2002) 203–222.

[6] T. Taniguchi, S. Koizumi, K. Watanabe, I. Sakaguchi, T. Sekiguchi, S. Yamaoka, High pressure synthesis of UV-light emitting cubic boron nitride single crystals, Diam Relat Mater. 12 (2003) 1098–1102.

[7] M.E. Turiansky, D. Wickramaratne, J.L. Lyons, C.G. Van De Walle, Prospects for n-type conductivity in cubic boron nitride, Appl Phys Lett. 119 (2021) 162105.

[8] O. Auciello, A. V. Sumant, Status review of the science and technology of ultrananocrystalline diamond (UNCD$^{TM}$) films and application to multifunctional devices, Diam Relat Mater. 19 (2010) 699–718.

[9] W. Zhang, Y.M. Chong, B. He, I. Bello, S-T. Lee, Cubic boron nitride: Properties and applications, Comprehensive Hard Metrials 3 (2014) 607–639.

[10] N. Izyumskaya, D.O. Demchenko, S. Das, Ü. Özgür, V. Avrutin, H. Morkoç, Recent development of boron nitride towards electronic applications, Adv. Electron. Mater. 3 (2017) 1600485.

[11] R.M. Chrenko, Ultraviolet and infrared spectra of cubic boron nitride, Solid State Commun. 14 (1974) 511–515.

[12] K. Chen, B. Song, N.K. Ravichandran, Q. Zheng, X. Chen, H. Lee, H. Sun, S. Li, G.A.G.U. Gamage, F. Tian, Z. Ding, Q. Song, A. Rai, H. Wu, P. Koirala, A.J. Schmidt, K. Watanabe, B. Lv, Z. Ren, L. Shi, D.G. Cahill, T. Taniguchi, D. Broido, G. Chen, Ultrahigh thermal conductivity in isotope-enriched cubic boron nitride, Science. 367 (2020) 555–559.

[13] C.B. Samantaray, R.N. Singh, Review of synthesis and properties of cubic boron nitride (c-BN) thin films, International Materials Reviews. 50 (2005) 313–344.

[14] X. Zhang, J. Meng, Recent progress of boron nitrides, Ultra-Wide Bandgap Semiconductor Materials. (2019) 347–419.

[15] P.B. Mirkarimi, K.F. McCarty, D.L. Medlin, Review of advances in cubic boron nitride film synthesis, Materials Science and Engineering R: Reports. R21 (1997) 47–100.

[16] X.W. Zhang, H.G. Boyen, H. Yin, P. Ziemann, F. Banhart, Microstructure of the intermediate turbostratic boron nitride layer, Diam Relat Mater. 14 (2005) 1474–1481.

[17] B. Zhong, T. Zhang, X.X. Huang, G.W. Wen, J.W. Chen, C.J. Wang, Y.D. Huang, Fabrication and Raman scattering behavior of novel turbostratic BN thin films, Mater Lett. 151 (2015) 130–133.

[18] T. Jähnichen, J. Hojak, C. Bläker, C. Pasel, V. Mauer, V. Zittel, R. Denecke, D. Bathen, D. Enke, Synthesis of Turbostratic Boron Nitride: Effect of Urea Decomposition, ACS Omega. 7 (2022) 33375–33384.

[19] P.B. Mirkarimi, ; K F Mccarty, ; G F Cardinale, ; D L Medlin, ; D K Ottesen, H.A. Johnsen, K.F. Mccarty, G.F. Cardinale, ) D L Medlin, D.K. Ottesen, Substrate effects in cubic boron nitride film formation, Journal of Vacuum Science & Technology A. 14 (1996) 251–255.

[20] W.J. Zhang, I. Bello, Y. Lifshitz, K.M. Chan, Y. Wu, C.Y. Chan, X.M. Meng, S.T. Lee, Thick and adherent cubic boron nitride films grown on diamond interlayers by fluorine-assisted chemical vapor deposition, Appl Phys Lett. 85 (2004) 1344–1346.

[21] S. Khan, F. Angeles, J. Wright, S. Vishwakarma, V.H. Ortiz, E. Guzman, F. Kargar, A.A. Balandin, D.J. Smith, D. Jena, H.G. Xing, R. Wilson, Properties for Thermally Conductive Interfaces with Wide Band Gap Materials, Applied Materials and Interfaces. 14 (2022) 36178–36188.

[22] X. Huang, Z. Guo, High thermal conductance across c-BN/diamond interface, Diam Relat Mater. 108 (2020) 107979.

[23] W.A. Little, The transport of heat between dissimilar solids at low temperatures, Can. J. Phys. 37 (1959) 334–349.

[24] E.T. Swartz, R.O. Pohl, Thermal boundary resistance, Rev Mod Phys. 61 (1989) 605–668.

[25] T. Feng, H. Zhou, Z. Cheng, L.S. Larkin, M.R. Neupane, A Critical Review of Thermal Boundary Conductance across Wide and Ultrawide Bandgap Semiconductor Interfaces, ACS Appl Mater Interfaces. 15 (2023) 29655–29673.

[26] F. Kargar, B. Debnath, J.-P. Kakko, A. Saÿnätjoki, H. Lipsanen, D.L. Nika, R.K. Lake, A.A. Balandin, Direct observation of confined acoustic phonon polarization branches in free-standing semiconductor nanowires, Nat Commun. 7 (2016) 13400.

[27] A.A. Balandin, D.L. Nika, Phononics in low-dimensional materials, Materials Today. 15 (2012) 266–275.

[28] M. Morita, T. Shiga, Surface phonons limit heat conduction in thin films, Phys Rev B. 103 (2021) 195418.

[29] R. Qi, R. Shi, Y. Li, Y. Sun, M. Wu, N. Li, J. Du, K. Liu, C. Chen, J. Chen, F. Wang, D. Yu, E.G. Wang, P. Gao, Measuring phonon dispersion at an interface, Nature. 599 (2021) 399–403.

[30] G. Chen, Nanoscale energy transport and conversion: a parallel treatment of electrons, molecules, phonons, and photons, Oxford University Press, 2005.

[31] Z. Cheng, R. Li, X. Yan, G. Jernigan, J. Shi, M.E. Liao, N.J. Hines, C.A. Gadre, J.C. Idrobo, E. Lee, K.D. Hobart, M.S. Goorsky, X. Pan, T. Luo, S. Graham, Experimental observation of localized interfacial phonon modes, Nature Communications 2021 12:1. 12 (2021) 1–10.

[32] F. Kargar, A.A. Balandin, Advances in Brillouin–Mandelstam light-scattering spectroscopy, Nat Photonics. 15 (2021) 720–731.

[33] J.R. Sandercock, Trends in brillouin scattering: Studies of opaque materials, supported films, and central modes, in: M. Cardona, G. Güntherodt (Eds.), Light Scattering in Solids III. Topics in Applied Physics, Springer, Berlin, Heidelberg, 1982: pp. 173–206.

[34] P. Mutti, C.E. Bottani, G. Ghislotti, M. Beghi, G.A.D. Briggs, J.R. Sandercock, Surface Brillouin scattering—Extending surface wave measurements to 20 GHz, in: A. Briggs (Ed.), Advances in Acoustic Microscopy, Springer, Boston, MA, 1995: pp. 249–300.

[35] Z.E. Nataj, Y. Xu, J. Brown, J. Garg, X. Chen, F. Kargar, A.A. Balandin, Cryogenic Characteristics of Graphene Composites - Evolution from Thermal Conductors to Thermal Insulators, Nature Communications 2023 14 (2023) 3190.

[36] S. Vishwakarma, J.M. Brown, A. Patel, M.R. McCartney, R.J. Nemanich, D.J. Smith, Growth and Characterization of Boron Nitride/Diamond Heterostructures, Microscopy and Microanalysis. 28 (2022) 2830–2831.

[37] E. Guzman, F. Kargar, F. Angeles, R.V. Meidanshahi, T. Grotjohn, A. Hardy, M. Muehle, R.B. Wilson, S.M. Goodnick, A.A. Balandin, Effects of Boron Doping on the Bulk and Surface Acoustic Phonons in Single-Crystal Diamond, ACS Appl Mater Interfaces. 14 (2022) 42223–42231.

[38] S. Prawer, R.J. Nemanich, Raman spectroscopy of diamond and doped diamond, Philosophical Transactions of the Royal Society of London. Series A: Mathematical, Physical and Engineering Sciences. 362 (2004) 2537–2565.

[39] V. Mortet, Z.V. Živcová, A. Taylor, M. Davydová, O. Frank, P. Hubík, J. Lorincik, M. Aleshin, Determination of atomic boron concentration in heavily boron-doped diamond by Raman spectroscopy, Diam Relat Mater. 93 (2019) 54–58.

[40] V. Mortet, I. Gregora, A. Taylor, N. Lambert, P. Ashcheulov, Z. Gedeonova, P. Hubik, New perspectives for heavily boron-doped diamond Raman spectrum analysis, Carbon. 168 (2020) 319–327.

[41] M. Mermoux, B. Marcus, G.M. Swain, J.E. Butler, A confocal Raman imaging study of an optically transparent boron-doped diamond electrode, Journal of Physical Chemistry B. 106 (2002) 10816–10827.

[42] A. Crisci, M. Mermoux, B. Saubat-Marcus, Deep ultra-violet Raman imaging of CVD boron-doped and non-doped diamond films, Diam Relat Mater. 17 (2008) 1207–1211.

[43] R.J. Nemanich, S.A. Solin, R.M. Martin, Light scattering study of boron nitride microcrystals, Phys Rev B. 23 (1981) 6348–6356.

[44] R. V Gorbachev, I. Riaz, R.R. Nair, R. Jalil, L. Britnell, B.D. Belle, E.W. Hill, K.S. Novoselov, K. Watanabe, T. Taniguchi, A.K. Geim, P. Blake, R. V Gorbachev, I. Riaz, R.R. Nair, R. Jalil, L. Britnell, B.D. Belle, E.W. Hill, K.S. Novoselov, A.K. Geim, P. Blake, K. Watanabe, T. Taniguchi, Hunting for Monolayer Boron Nitride: Optical and Raman Signatures, Small. 7 (2011) 465–468.

[45] S. Reich, A.C. Ferrari, R. Arenal, A. Loiseau, I. Bello, J. Robertson, Resonant Raman scattering in cubic and hexagonal boron nitride, Phys Rev B. 71 (2005).

[46] O. Kutsay, C. Yan, Y.M. Chong, Q. Ye, I. Bello, W.J. Zhang, J.A. Zapien, Z.F. Zhou, Y.K. Li, V. Garashchenko, A.G. Gontar, N. V. Novikov, S.T. Lee, Studying cubic boron nitride by Raman and infrared spectroscopies, Diam Relat Mater. 19 (2010) 968–971.

[47] K. Luo, Y. Zhang, D. Yu, B. Li, W. Hu, Y. Liu, Y. Gao, B. Wen, A. Nie, Z. Zhao, B. Xu, X.F. Zhou, Y. Tian, J. He, Small onion-like BN leads to ultrafine-twinned cubic BN, Sci China Mater. 62 (2019) 1169–1176.

[48] M.I. Eremets, M. Gauthier, A. Polian, J.C. Chervin, J.M. Besson, G.A. Dubitskii, Y.Y. Semenova, Optical properties of cubic boron nitride, Phys Rev B. 52 (1995) 8854–8863.

[49] O. Stenzel, J. Hahn, M. Röder, A. Ehrlich, S. Prause, F. Richter, The Optical Constants of Cubic and Hexagonal Boron Nitride Thin Films and Their Relation to the Bulk Optical Constants, Physica Status Solidi (a). 158 (1996) 281–287.

[50] X. Jiang, J. v. Harzer, B. Hillebrands, Ch. Wild, P. Koidl, Brillouin light scattering on chemical-vapor-deposited polycrystalline diamond: Evaluation of the elastic moduli, Appl Phys Lett. 59 (1991) 1055–1057.

[51] R.J. Jiménez-Riobóo, L. Artús, R. Cuscó, T. Taniguchi, G. Cassabois, B. Gil, In- and out-of-plane longitudinal acoustic-wave velocities and elastic moduli in h-BN from Brillouin scattering measurements, Appl Phys Lett. 112 (2018) 051905.

[52] G. Carlotti, Elastic Characterization of Transparent and Opaque Films, Multilayers and Acoustic Resonators by Surface Brillouin Scattering: A Review, Applied Sciences 2018, Vol. 8, Page 124. 8 (2018) 124.

[53] G.W. Farnell, E.L. Adler, Elastic Wave Propagation in Thin Layers, Physical Acoustics. 9 (1972) 35–127.

[54] J. Du, B.R. Tittmann, H.S. Ju, Evaluation of film adhesion to substrates by means of surface acoustic wave dispersion, Thin Solid Films. 518 (2010) 5786–5795.

[55] H. Xia, G.X. Cheng, G.G. Liu, W. Zhang, K.J. Chen, X.K. Zhang, Brillouin scattering study of interface constituents in a-Si:H/a-SiNx:H superlattices, Solid State Commun. 73 (1990) 657–660.